\documentclass[A4,paper,12pt]{article}

\usepackage{amsmath}
\usepackage{amssymb}
\usepackage[cp1250]{inputenc}
\usepackage{placeins}
\usepackage{graphicx}

\renewcommand{\d}{\mathsf{d}}
\newcommand{\opr}[1]{\mathsf{#1}}
\newcommand{\form}[1]{\mathsf{#1}}
\newcommand{\sobsp}{\mathcal{H}}

\renewcommand{\[}{\begin{equation}}                               %
\renewcommand{\]}[1]{\label{eq:#1}\end{equation}}

\newcommand{\laplace}{\triangle}
\newcommand{\grad}{\bigtriangledown}

\newenvironment{proof}{{\sl Proof:}\rm}{$\blacksquare$ \\ \smallskip}

\newtheorem{defin}{Definition}[section]
\newtheorem{lemma}[defin]{Lemma}
\newtheorem{teorem}[defin]{Theorem}

\newtheorem{apropos}[defin]{Proposition}

\begin{document}

\begin{flushleft}

\textbf{\Large On the dense point and absolutely continuous \\
[0em] spectrum for Hamiltonians with concentric $\delta$ \\
[.3em] shells} \\
[2em]
{\large Pavel Exner and Martin Fraas} \\ [.2em]
{\small \emph{Nuclear Physics Institute, Czech Academy of
Sciences,
25068 \v{R}e\v{z} near Prague, \\
Doppler Institute, Czech Technical University, B\v{r}ehov\'{a} 7,
11519 Prague, Czechia \\ e-mail: exner@ujf.cas.cz,
fraam0am@artax.karlin.mff.cuni.cz}
\\ [1em]
{\small \textbf{Abstract.} We consider Schr\"odinger operator in
dimension $d\ge 2$ with a singular interaction supported by an
infinite family of concentric spheres, analogous to a system
studied by Hempel and coauthors for regular potentials. The
essential spectrum covers a halfline determined by the appropriate
one-dimensional comparison operator; it is dense pure point in the
gaps of the latter. If the interaction is radially periodic, there
are absolutely continuous bands; in contrast to the regular case
the measure of the p.p. segments does not vanish in the
high-energy limit.}
\\ [1.5em]
\textbf{Mathematics Subject Clasification (2000)}. 35J10, 35P99,
81Q10
\\ [.5em]
 }
\end{flushleft}

\vspace{3em}

\setcounter{equation}{0}
\section{Introduction}

Operators the spectrum of which consists of interlaced components
of different spectral types are always of interest. One of the
situations where they can occur concerns radially symmetric and
periodic potentials.

The idea can be traced back to the paper \cite{Hempel} by Hempel,
Hinz, and Kalf who asked whether the gaps in the spectrum of the
one-dimensional Schr\"{o}dinger operator
\[- \frac{\d^2}{\d r^2} + q(r) ,\]{2.0.1}
with an even potential, $q(-r) = q(r)$, are preserved or filled up
as one passes to the spherically symmetric operator
\[ -\laplace + q(|\cdot|) \quad \mbox{in} \quad
L^2(\mathbb{R}^\nu),\quad \nu \geq 2 .\]{2.0.2}
They proved that for a potential which not oscillate too rapidly and
belongs to $L^1_\mathrm{loc}(\mathbb{R})$, the negative part having
this property uniformly, the gaps are filled, i.e. the essential
spectrum covers the half-line $[\lambda_0,\,\infty)$, where
$\lambda_0$ is the essential-spectrum threshold of the associated
one-dimensional operator (\ref{eq:2.0.1}). In the subsequent paper
\cite{Hempel2} Hempel, Herbst, Hinz, and Kalf proved that if $q$ is
periodic on the half-line the absolutely continues spectra is
preserved and the gaps are filled with a dense point spectrum.

The spectrum of such systems has been studied further from the
viewpoint of the eigenvalue distribution in the gaps \cite{Brown}
and it was also show that the system has a family of \emph{isolated}
eigenvalues accumulating at the essential-spectrum threshold
\cite{Brown2}. An extension to magnetic Schr\"odinger operators
\cite{Hoever} and Dirac operators \cite{Schmidt} were also
considered.

A characteristic property of such an interlaced spectrum is that
the intervals of the dense pure point spectrum shrink as the
energy increases. The aim of this letter is to present an example
where the width of the dense-point ``bands'' \emph{remains
nonzero} in the high-energy limit. Since the asymptotic behavior
is determined by that of the underlying one-dimensional problem,
and thus by the regularity of the potential $q$, it is clear that
we have choose a singular one; we will investigate a family of
Schr\"odinger operators given formally by
 $$
\opr{H} = -\laplace + \alpha \sum\limits_n \delta(|x| - R_n)\quad
\mbox{in} \quad L^2(\mathbb{R}^\nu),\quad \nu \geq 2,
 $$
with a $\delta$ interaction supported by a family of concentric
spheres. We will describe the model properly in the next section,
then we determine its essential spectrum, and in
Section~\ref{s:structure} we will show the indicated spectral
property.

\setcounter{equation}{0}
\section{Description of the model}
\label{s:kropen}

Let us first briefly recall properties of the one-dimensional
systems with $\delta$ interactions \cite{Solvable}. The operator
$\opr{h} = -\laplace + \alpha \sum_{n \in \mathbb{Z}} \delta(x -
x_n)$ can be given meaning if we require that the points
supporting the interaction do not accumulate, $\inf|x_n - x_m|>0$.
Then one can check that the symmetric form $\form{t_\alpha}$
defined by
\[
\form{t_\alpha}[f,\,g] = (f',\,g') + \alpha \sum\limits_{n \in
\mathbb{Z}} f(x_n)\bar{g}(x_n),\quad
 D(\form{t_\alpha}) = \sobsp^{1,\,2}(\mathbb{R}),\]{m1}
is closed and bounded from below \cite{Solvable, Brasche}, and we
identify the corresponding self-adjoint operator $\opr{h}_\alpha$,
in the sense of first representation theorem \cite{Kato} with the
formal operator mentioned above. One can describe it explicitly in
terms of boundary conditions: it acts as $\opr{h}_\alpha f = -f''$
on the domain
  $$
  D(\opr{h}_\alpha) = \left\{\, f \in \sobsp^{2,\,2}\Big(\mathbb{R}
  \backslash \bigcup\limits_{n \in \mathbb{Z}} \{x_n\} \Big):\:
  f'(x_n+) - f'(x_n-) = \alpha f(x_n)\right\} .
  $$
The Kronig-Penney model corresponds to a periodic arrangement of
the $\delta$-interactions, for instance, $x_n = \left(n-\frac12
\right)a$ for some $a > 0$. It has a purely absolutely continuous
spectrum with the known band structure \cite{Solvable} and these
properties do not change when we pass to such a system on a
half-line with any boundary condition at the origin, the only
change is that the spectral multiplicity will be one instead of
two.

After this preliminary let us pass to our proper topic and define
an operator which can be identified with (\ref{eq:2.0.2}); we
suppose again that the sequence of radii can accumulate only at
infinity, $\inf|R_n - R_m|>0$. As above we employ the appropriate
symmetric form
$$
\form{T}_\alpha[f,\,g] = \int\limits_{\mathbb{R}^n}\grad f(x)
\cdot \grad \bar{g}(x)\,\d^nx + \alpha \sum_n
\int\limits_{S_{R_n}} f(x)\bar{g}(x)\,\d\Omega,
$$
with $D(\form{T}_\alpha) = \sobsp^{1,\,2}(\mathbb{R}^n)$, where
$S_{R_n}$ is the sphere of radius $R_n$ and $d\Omega$ is the
corresponding ``area'' element. Since the form is spherically
symmetric, it is natural to use a partial wave decomposition.
Consider the isometry
$$
    \mathsf{U}\,:\,L^2((0,\,\infty),r^{\nu-1}dr)\, \rightarrow
    \, L^2(0,\,\infty), \quad \mathsf{U}f(r) =
    r^{\frac{\nu-1}{2}}f(r),
$$
which allows us to write
$$
    L^2(\mathbb{R}^\nu) = \bigoplus_l\,
    \mathsf{U}^{-1}L^2(0,\,\infty)\otimes L^2(S_1)
$$
and
$$
   \form{T}_\alpha = \bigoplus_l\, \mathsf{U}^{-1}
   \form{T}_{\alpha,\,l}\mathsf{U} \otimes \opr{I}_l ,
$$
where $\opr{I}_l$ is the unit operator on $L^2(S_1)$ and
    \begin{multline}
    \form{T}_{\alpha,\,l}[f,g]= \\ \int\limits_0^\infty
    \left( f'(r)\bar{g}'(r) + \frac{1}{r^2}
    \left[\frac{(n-1)(n-3)}{4} + l(l+n-2)\right]f(r)
    \bar{g}(r) \right)\d r \\
    + \alpha \sum_n f(R_n)\bar{g}(R_n),
     \nonumber
    \end{multline}
with $D(\form{T}_{\alpha,\,l})=\sobsp^{1,\,2}(0,\infty)$. The
following lemma will help us to find properties of the form
$\form{T}_{\alpha,\,l}$.

\begin{lemma}
\begin{enumerate}
\item[(i)] Let $a>0$. There exists a positive $ b $ so that
\[ |\alpha| \sum_n|f(R_n)|^2 \leq a \int\limits_0^\infty |f'(x)|^2 \d x + b
\int\limits_0^\infty |f(x)|^2 \d x \]{2.8}
holds for all functions $f$ belonging to the Schwartz space $
\mathcal{S}(0,\,\infty)$.
 \item[(ii)] There exist $C$ such that, for every function $f$ in
 the domain of $ \opr{H}_{\alpha,\,l} $
holds\footnote{The operator $\opr{H}_{\alpha,\,l}$ associated with
$\form{T}_{\alpha,\,l}$ is described explicitly Theorem~\ref{pwd}
below.}
\[||f'|| \leq C(||\opr{H}_{\alpha,\,l}f|| + ||f||) \]{2.9}
\end{enumerate}
\label{def:pom}
\end{lemma}
\begin{proof}
Let $I \subset \mathbb{R}_+$ be an interval and $f \in
\sobsp^{1,\,2}(I)$. By a standard embedding we have
$\sobsp^{1,\,2}(I) \hookrightarrow \mathcal{C}(I)$, more
explicitly, there is a $C>0$ such that
\[
  |f(x)|^2 \leq C \left(\int_I |f(y)|^2\, \d y
  + \int_I |f'(y)|^2\, \d y \right)
\]{embedding}
holds for every $x \in I$. Let $\{y_n\}_{n=0}^\infty$ be an
increasing sequence of positive numbers such that $\sup|y_{n+1} -
y_n|>2$ and $y_1 \ge 1$. Then we consider the family of mutually
disjoint intervals $I_n = (y_n -1,\,y_n+1)$ and summing the
inequalities (\ref{eq:embedding}) for $I=I_n$ over $n$ we get
$$
\sum_{n=1}^\infty |f(y_n)|^2 \leq C \left(\int_0^\infty |f(y)|^2\,
\d y + \int_0^\infty |f'(y)|^2\, \d y \right).
$$
To conclude the argument we employ a scaling. The last inequality
applied to  $f_\varepsilon:\: f_\varepsilon(x)=f(\varepsilon x)$
gives
$$
\sum_{n=1}^\infty |f(\varepsilon y_n)|^2 \leq C
\left(\varepsilon^{-1} \int_0^\infty |f(y)|^2\, \d y +
\varepsilon\, \int_0^\infty |f'(y)|^2\, \d y \right);
$$
the claim (i) then follows by substitution
$y_n=R_n\,\varepsilon^{-1}$ with $\varepsilon$ such that $C
\varepsilon < a|\alpha|^{-1}$ and $\sup|R_{n+1} -
R_n|>2\varepsilon$, since without loss of generality we may
suppose that $\alpha\ne 0$. The claim (ii) in turn follows from
(i) with a fixed $a < 1$ together with the inequality
\begin{multline}
 ||f'||^2 = (\opr{H}_{\alpha,\,l}f,f)
 - \int\limits_0^\infty \frac{1}{r^2}\left(\frac{(n-1)(n-3)}{4} +
l(l+n-2)\right)|f(r)|^2\, \d r \\ - \alpha \sum_n|f(R_n)|^2  \leq
\frac{1}{2}||\opr{H}_{\alpha,\,l}f||^2 + \frac{1}{2}||f||^2
+a||f'||^2 + b||f||^2, \nonumber \end{multline}
where we used Cauchy-Schwarz inequality,
$(\opr{H}_{\alpha,\,l}f,f) \leq \frac{1}{2}
(||\opr{H}_{\alpha,\,l}f||^2 + ||f||^2)$, and the nonnegativity of
the second term.
\end{proof}

This allows us to describe the model Hamiltonian explicitly in
terms of boundary conditions at the singular points.

\begin{teorem} \label{pwd}
\begin{enumerate}
\item[(i)] The quadratic form $ \form{T}_{\alpha,\,l}$ is bounded
from below and closed on $L^2(0,\,\infty)$ and the space $
C_0^\infty(0,\,\infty) $ of infinitely differentiable functions of
compact support is a core of $ \form{T}_{\alpha,\,l} $.
 \item[(ii)] The self-adjoint operator corresponding to
$\form{T}_{\alpha,\,l}$ by the first representation theorem is
$$
\opr{H}_{\alpha,\,l} = -\frac{\d^2}{\d^2 r} +
\frac{1}{r^2}\left(\frac{(n-1)(n-3)}{4} + l(l+n-2)\right),
$$
with the domain $D(\opr{H}_{\alpha,\,l})$ given by
\begin{multline}
\left\{ f\in \sobsp^{2,\,2} \left(\mathbb{R}^{+} \setminus
\bigcup\limits_{n}\, \{R_n\} \right):\: f'(R_n+) - f'(R_n-) =
\alpha f(R_n) \right\},\nonumber
\end{multline}
and the self-adjoint operator associated with the
$\form{T}_\alpha$ is thus
  \[
   \opr{H}_\alpha = \bigoplus_l\: \mathsf{U}^{-1}
   \opr{H}_{\alpha,\,l}\mathsf{U} \otimes
   \opr{I}_l.
  \]{oprdef}
\end{enumerate}
\end{teorem}
\begin{proof}
The first claim follows from Ref.~\cite{Brasche} in combination with
the previous lemma, the second one can be verified directly.
\end{proof}

\setcounter{equation}{0}
\section{The essential spectrum} \label{s:4}

Let us first introduce some notation which we will use throughout
this section. We need a one-dimensional comparison operator. For
simplicity we take an operator on the whole axis extending the
family $\{R_n\}_{n=1}^\infty$ of the radii to $\{R_n\}_{n\in
\mathbb{Z}}$ by putting $R_{-n} = -R_{n+1}$ for $n=0,1,\dots$. By
$\opr{h_{\alpha}}$ we denote the self-adjoint operator defined in
the opening of the previous section in which we now put
$x_n:=R_n$; the corresponding quadratic form will be again denoted
as $\form{t}_\alpha $. By $ \opr{h}_{\alpha,\,R} $ we denote the
self-adjoint operator obtained from $\opr{h}_\alpha$ by adding the
Dirichlet boundary conditions at the points $\pm R$. Since
$\opr{h}_\alpha$ and $\opr{h}_{\alpha,\,R}$ have a common
symmetric restriction with finite deficiency indices we have
\[\sigma_{ess}(\opr{h}_\alpha)
= \sigma_{ess}(\opr{h}_{\alpha,\,R}). \]{3.2.0}
Furthermore, by $\opr{h}_{\alpha,\,(a,\,b)}$ and
$\opr{h}_{\alpha,\,R,\,(a,\,b)}$ we denote the self-adjoint
operator which is a restriction of $\opr{h}_\alpha$,
$\opr{h}_{\alpha,\,R}$ to $L^2(a,\,b)$, respectively, with
Dirichlet boundary conditions at the interval endpoints. We note
that
\[
 \opr{h}_{\alpha,\,R,\,(0,\,\infty)} = \opr{h}_{\alpha,\,(0,\,R)}
\oplus \opr{h}_{\alpha,\,(R,\,\infty)}.
\]{3.2.01}
We use a similar notation, namely $\opr{H}_{\alpha,\,l,\,R}$ and
$\opr{H}_{\alpha,\,l,\,(a,\,b)}$, for operators in every partial
wave. Furthermore $\opr{H}_{\alpha,\,(\rho,\,R)}$ denotes the
restriction of $\opr{H}_\alpha$ to the spherical shell $B_R
\setminus B_\rho$. Our main result in this section reads as
follows.
\begin{teorem}{The essential spectrum of the operator
(\ref{eq:oprdef}) is equal to}
\[\sigma_{ess}(\opr{H}_\alpha) =
[\inf \sigma_{ess}(\opr{h}_\alpha),\infty)\]{3.2.1}
\label{def:ess}
\end{teorem}
The idea of the proof is the same as in \cite{Hempel}. First we
check that $\inf \sigma_{ess}(\opr{H}_\alpha)$ cannot be smaller
then $\inf \sigma_{ess}(\opr{h}_\alpha)$, after that we will show
that $\sigma_{ess}(\opr{H}_\alpha)$ contains the interval $[\inf
\sigma_{ess}(\opr{h}_\alpha),\infty)$.
\begin{apropos}{In the stated assumptions we have}
\[\inf \sigma_{ess}(\opr{H}_\alpha) \geq
\inf \sigma_{ess}(\opr{h}_\alpha) \]{3.2.2}
\label{def:p1}
\end{apropos}
\begin{proof}
The partial-wave decomposition of Theorem~\ref{pwd} in combination
with the minimax principle imply that the spectral minimum is
reached in the $s$-state subspace, hence we can consider only
spherically symmetric functions. Then the idea is to estimate $\inf
\sigma_{ess}(\opr{H}_\alpha)$ by means of the lowest eigenvalue
$\mu_{\rho,\,R}$ of the operator $\ \opr{H}_{\alpha,\,(\rho,\,R)}$
and $\rho,\,R$ large enough. The associated -- spherically symmetric
-- eigenfunction $u_{\rho\,R}$ clearly satisfied the $\delta$
boundary conditions, hence one can repeat the argument from
\cite{Hempel}, Proposition~1.
\end{proof}

\begin{apropos}
\[\sigma_{ess}(\opr{H}_\alpha) \supset [\inf \sigma_{ess}(\opr{h}_\alpha),\infty)
\]{3.2.3}
\label{def:p2}
\end{apropos}
\begin{proof}
The idea is to employ Weyl criterion. Following \cite{Weid}, let
$\lambda_0 \in \sigma_{ess}(h_\alpha)$ and $ \lambda
> 0 $, then we have to show that for every $\epsilon >0$ there is a
function
$$
 \varphi \in D(\opr{H}_{\alpha})\quad \mbox{satisfying}\quad
||\varphi|| \geq 1\quad \mbox{and}\quad||(\opr{H}_{\alpha} -
\lambda_0 - \lambda)\varphi|| \leq \epsilon .
$$
The key ingredients in the estimates of the regular-case proof --
cf.~\cite{Weid}, (i), (ii) on the first page -- correspond to the
equations (\ref{eq:2.9}) and (\ref{eq:3.2.0}) here. In order to use
directly the said argument, we have to deal with the boundary
conditions. To do this we use the simple observation that whenever
\[
f(r) \in D(\opr{h_{\alpha}}) \quad \mbox{and} \quad g(x) \in
D(\opr{H}_0) \quad \mbox{then} \quad \phi(x) = f(|x|)g(x) \in
D(\opr{H}_\alpha)\,,
\]{3.2.5}
now we consider such a $\phi(x)$ and follow step by step the proof
in \cite{Weid}.
\end{proof}

\setcounter{equation}{0}
\section{Character of the spectrum}
\label{s:structure}

In this section we will make two claims. One is general, without a
specific requirement on the distribution of the $\delta$ barriers
other that $\inf|R_n - R_m|>0$. It stems from the fact that the
essential spectrum of the associated one-dimensional operator
$\opr{h}_\alpha$ may have gaps; we want to know how the spectrum
of $\opr{H_\alpha}$ looks like in these gaps. First we observe
that in every partial wave
\[
\sigma_{ess}(\opr{H}_{\alpha,\,l}) = \sigma_{ess}
(\opr{h}_\alpha) .
\]{3.2.16}
Indeed, in view of (\ref{eq:3.2.0}) we have
$$
\sigma_{ess}(\opr{H}_{\alpha,\,l}) =
\sigma_{ess}(\opr{H}_{\alpha,\,l,\,R})\,,
$$
and since $\opr{H}_{\alpha,\,l,\,(0,\,R)}$ has a purely discrete
spectrum, we use (\ref{eq:3.2.01}) to infer that
\[
 \sigma_{ess}(\opr{H}_{\alpha,\,l}) =
\sigma_{ess}(\opr{H}_{\alpha,\,l,\,(R,\,\infty)})\,.
\]{3.2.19}
Furthermore, a multiplication by (a multiple of) $r^{-2} $ is $
\opr{h}_{\alpha,\,(R,\,\infty)} $ compact, which implies by Weyl's
theorem that
$$
\sigma_{ess}(\opr{H}_{\alpha,\,l,(R,\,\infty)}) =
\sigma_{ess}(\opr{h}_{\alpha,\,(R,\,\infty)})\,,
$$
and using once more the ``chopping'' argument we arrive at
(\ref{eq:3.2.16}). Now we are ready to state and prove the claim
which is a counterpart of the result derived in \cite{Hempel2} for
regular potential barriers.

\begin{teorem}\label{def:summarize}
Let $\opr{H}_\alpha$ be as described above, then for any gap
$(\alpha,\,\beta)$ in the essential spectrum of $\opr{h}_\alpha $
the following is valid:
\begin{description}
\item (i) $\opr{H}_\alpha$ has no continuous spectrum in $
(\alpha,\,\beta) $; \item (ii) eigenvalues of $\opr{H}_{\alpha}$
are dense in $(\alpha,\,\beta)$.
\end{description}
\label{def:main}
\end{teorem}
\begin{proof}
By (\ref{eq:3.2.16}), none of the operators $\opr{H}_{\alpha,\,l},\,
l=0,\,1,\,2,\dots$, has a continuous spectrum in $(\alpha,\,\beta)$,
hence $\opr{H}_{\alpha}$ has no continuous spectrum in this interval
either. On the other hand, the entire interval $(\alpha,\,\beta)$ is
contained in the essential spectrum of $\opr{H}_\alpha$, and it
follows that the spectrum of $\opr{H}_\alpha$ in $(\alpha,\,\beta)$
consists of eigenvalues, which are necessarily dense in the
interval.
\end{proof}

Now we pass to a particular case when the $\delta$-sphere
interactions are arranged in a periodic way, $R_n = na - a/2$ with
\mbox{$a>0$}, and prove that in this situation there is a purely
continuous spectrum in the \emph{bands} of the associated
one-dimensional Kronig-Penney model. The argument is similar to
Section~2 of \cite{Hempel2} so we will concentrate mostly on the
changes required by the singular character of the interaction.

\begin{lemma}
Let $(a,\,b)$ be the interior of a band of the operator
$\opr{h}_\alpha$ in $L_2(\mathbb{R})$. Let further $K \subset
(a,\,b)$ be a compact subinterval, $c \in \mathbb{R}$, and $x_0 >
0$. Then there exist numbers $C_1,\,C_2 >0$ such that for every
$\lambda \in K$ any solution $u$ of
\[
-u''(r) + \frac{c}{r^2}u(r) = \lambda u(r)\,, \quad u \in
D(\opr{h}_\alpha)\,,
\]{pur.1}
with the normalization $|u(x_0)|^2 + |u'(x_0)|^2 = 1$ satisfies
\[
C_1^2 \geq |u(x)|^2 + |u'(x)|^2\,, \quad \int\limits_{x_0}^{x}
|u(t)|^2\,\d t \geq C_2(x - x_0) \quad\mathrm{for}\;\; x \geq x_0
+1\,.
\]{pur.3}
\label{def:pur.1}
\end{lemma}
\begin{proof}
Let $\lambda \in K$. As it is well known \cite{Weid2} the equation
$\opr{h}_\alpha w = \lambda w$ has two linearly independent
solutions $ u_0 = u_0(\cdot,\,\lambda),\,v_0 =
v_0(\cdot,\,\lambda)$ such that $u_0,\,v_0 \in D(\opr{h}_\alpha)$,
and $|u_0|,\,|u'_0|, \,|v_0|,\,|v'_0| $ are periodic, bounded and
continuous w.r.t. $\lambda$. Without loss of generality we may
assume that the Wronski matrix
  $$
   Y = \left[\begin{matrix}
                u_0 & v_0 \\
                u'_0 & v'_0
        \end{matrix}\right]
   $$
has determinant equal to one. Let $C_0 > 0$ be a constant such
that
$$
|u_0(x,\,\lambda)|^2 + |u'_0(x,\,\lambda)|^2 +
|v_0(x,\,\lambda)|^2 + |v'_0(x,\,\lambda)|^2\leq C_0 \quad (x\in
\mathbb{R},\,\lambda \in K).
$$
Given any solution $ u $ of (\ref{eq:pur.1}), the function
$$
y := Y^{-1} \left[ \begin{matrix}
                        u \\
                        u'
                    \end{matrix} \right]
$$
satisfies the equation $y'= A y$ on every interval
$\left((n-\frac12)a,\,(n+\frac12)a\right)$, where
$$
A = -\frac{c}{x^2} \left[ \begin{matrix}
                                u_0 v_0 & v_0^2 \\
                                -u_0^2 & -u_0v_0
                            \end{matrix} \right]
$$
in analogy with \cite{Hempel2}. By a straightforward calculation
we get
$$
y = \left[\begin{matrix}
                v'_0 u - v_0 u' \\
                -u'_0 u + u_0 u'
            \end{matrix} \right]\,, \quad
y' = \frac{c}{x^2} \left[ \begin{matrix}
                                -v_0 u \\
                                u_0 u
                            \end{matrix} \right]\,,
$$
which implies that $y,\,y'$ are continuous at the singular points.
Thus
$$
y(x) = \exp\left\{\int\limits_{x_0}^x A(t)\, \d t\right\} y(x_0)
$$
is a solution of $y'= A y$ and as in \cite{Hempel2} it holds that
$$
\frac{1}{2}(|y|^2)'\leq |(y,\,y')| \leq \|A\||y|^2
$$
and so for $x \geq x_0$ we have
$$
|y(x)|^2 \leq |y(x_0)|^2 \exp\left\{2 \int\limits_{x_0}^x \|A(t)\|
\,\d t \right\} \leq |Y^{-1}(x_0)|^2 \exp\left\{ 2
\int\limits_{x_0}^\infty \|A(t)\| \,\d t\right\}
$$
for any solution of (\ref{eq:pur.1}) with the normalization
$|u(x_0)|^2 + |u'(x_0)|^2 = 1$. From
$$
\left[ \begin{matrix}
                u(x) \\
                u'(x)
            \end{matrix} \right] = Y(x) Y^{-1}(x_0) \left[ \begin{matrix}
                u(x_0) \\
                u'(x_0)
            \end{matrix} \right] + \int\limits_{x_0}^{x} Y(x) A(t) y(t)
            \,\d t\,, \quad x \geq x_0\,,
$$
we now infer the existence of a number $C_1 > 0$ such that
\[
|u(x)|^2 + |u'(x)|^2 \leq C_1^2\,, \quad x \geq x_0\,,
\]{pur.16}
holds for all solutions of (\ref{eq:pur.1}) which are normalized
in the described way. This proves the first inequality in
(\ref{eq:pur.3}).

Let $ u $ be a real-valued solution of (\ref{eq:pur.1}), again
with the same normalization, and suppose that $v$ is a solution
such that
$$
 v(x_0) = -u'(x_0)\,, \quad v'(x_0) = u(x_0)\,.
$$
Then the Wronskian of $ u $ and $ v $ equals one, and therefore
$$
1 = [u(x)v'(x) - u'(x) v(x)]^2 \leq [u^2(x) + u'^2(x)][v^2(x) +
v'^2(x)]\,, \quad x \geq x_0\,.
$$
Since $v$ satisfies (\ref{eq:pur.16}) we find that
$$
\frac{x-x_0}{C_1^2} \leq \int_{x_0}^{x}(u^2 + u'^2)(t)\, \d t\,,
\quad x \geq x_0\,,
$$
and the second assertion in (\ref{eq:pur.3}) follows from Lemma
\ref{def:pom}(ii)
\end{proof}

In particular, this lemma proves through (\ref{eq:pur.3}) that the
operator $\opr{H}_{\alpha,\,l}$ has no embedded eigenvalues in
$(a,\,b)$. Next we will derive a Lipschitz bound for the number of
eigenvalues of the operator $ \opr{h}_k \equiv
\opr{h}_{\alpha,\,(0,\,R_k + a/2)}$; we denote their number in the
interval $(\lambda_1,\,\lambda_2)$ by $N_k(\lambda_1,\, \lambda_2)$.

\begin{lemma}
  Let $(a,\,b)$ be a spectra band of the operator
  $\opr{h}_\alpha$ in $L^2(\mathbb{R})$ and $ \lambda_2 - \lambda_1>0$.
   Then there exists a number $ C >0 $ such that
  \[ N_k(\lambda_1,\,\lambda_2) \leq C(\lambda_2 - \lambda_1)R_k
  \]{pur.20}
  for every $ k \in \mathbb{N}$.
  \label{def:pur.2}
\end{lemma}
\begin{proof}
 Let $\opr{h}^{(\theta)}$ be the operator $\opr{h}_{\alpha}$
 acting on $L^2(0,\,a)$ with $\theta$-periodic boundary
 conditions. Then $\lambda$ is an eigenvalue of $\opr{h}_k$ if and
 only if there is an integer $j \in \{0,\,\dots,\,k-1\}$ such that $\lambda$ is the eigenvalue of
 $\opr{h}^{(j \pi /k)}$. The eigenvalues of $\opr{h}^{(\theta)}$ are
 the roots of Kronig-Penney equation,
 $$ \cos(\theta a) = \cos(\lambda a) + \frac{\alpha}{2 \lambda}
 \sin (\lambda a) \,.$$
It follows from Theorem~III.2.3.1 in \cite{Solvable} that there is
precisely one eigenvalue of $\opr{h}^{(\theta)}$ in every interval
$((k-1)^2 \pi^2 a^{-2},\,k^2 \pi^2 a^{-2})$. Hence
$$ N_k(\lambda_1,\,\lambda_2) \leq k \left\lceil (\sqrt{\lambda_2} -
\sqrt{\lambda_1}) \frac{a}{\pi} \right\rceil \leq
k\left((\sqrt{\lambda_2} - \sqrt{\lambda_1})\frac{a}{\pi} +
1\right) \leq R_k(\lambda_2 - \lambda_1) C\,,$$
where
 $$
 C:= 2\,\frac{a(\sqrt{\lambda_2} -
\sqrt{\lambda_1})+\pi}{a\pi(\lambda_2 - \lambda_1)}\,;
$$
we have used here the fact that $R_k-\frac12 a> \frac12 ka$.
 \end{proof}

With these preliminaries, we are prepared to prove the absolute
continuity of the spectrum inside the Kronig-Penney bands.

\begin{teorem}
The spectrum of $\opr{H}_{\alpha,\,l}$ is absolutely continuous in
the interior of each spectral band of $\opr{h}_\alpha$.
\end{teorem}
\begin{proof}
Since the argument is similar to \cite{Hempel2},
\cite[Thm~15.3]{Weid2}, we just sketch it. The aim is to show that
for any fixed $f \in C_0^\infty(0,\,\infty)$ the function $
||E(\lambda)f||^2$, where $E(\lambda)$ denotes the spectral
measure of $\opr{H}_{\alpha,\,l}$, is Lipschitz continuous for
$\lambda$ in the spectral band $(a,\,b)$. As there are no
eigenvalues of $\opr{H}_{\alpha,\,l}$ in $(a,\,b)$ by
Lemma~\ref{def:pur.1}) one has the strong convergence
$$
E^{R_n}(\lambda) \to E(\lambda)\,, \quad R_n \to \infty\,,
$$
where $E^{R_n}(\lambda)$ denotes the spectral resolution of
$\opr{H}_k := \opr{h}_k + c\,r^{-2}$, and consequently, it is
sufficient to prove that for $[\alpha,\,\beta] \subset (a,\,b)$
\[ ((E^{R_n}(\beta) - E^{R_n}(\alpha))f,\,f) \leq \mathrm{const}\,
(\beta - \alpha +\epsilon)\,. \]{pur.22}
holds for any $\epsilon$. The spectrum of
$\opr{H}_{\alpha,\,l,\,R_n}$ is purely discrete and simple. Let us
denote its $j$-th eigenvalue by $\lambda_j$ and suppose that the
associated eigenfunction $\phi_j$ has the normalization
$$ |\phi_j(R_0)|^2 + |\phi_j'(R_0)|^2 =1. $$
Lemma~\ref{def:pur.1} establishes the existence of numbers
$C_1,\,C_2>0$ such that
\begin{multline}
((E^{R_n}(\beta) - E^{R_n}(\alpha))f,\,f) \leq \sum_{\alpha <
\lambda_j <\beta} |(f,\,\phi_j)|^2||\phi_j||^{-2} \\ \leq
\frac{C_1^2}{C_2 (R_n-R_0)}||f||^2 \sum_{\alpha < \lambda_j<\beta}
1 \leq \frac{C_3}{R_n - R_0}\,\#\{j:\alpha < \lambda_j \leq
\beta\}\,, \label{eq:odhad}
\end{multline}
for all $R_n>R_0$. Now we fix $\varepsilon$ so small that $[\alpha
- \varepsilon/2,\,\beta + \varepsilon/2] \subset (a,\,b)$ and
choose $R_{n(\varepsilon)}$ so that
\[ \frac{|c|}{r^2} < \frac{\varepsilon}{2} \quad \mbox{for} \quad
r>R_{n(\varepsilon)}\]{Reps}
and impose an additional Dirichlet boundary condition at the point
$R_{n(\varepsilon)}$. Then the interval $(0,\,R_{n(\varepsilon)})$
contributes by a certain number $C_\varepsilon$ of eigenvalues. On
the other hand, from Lemma~\ref{def:pur.2} we know that the number
of eigenvalues of the operator $\opr{h}_{(k_\varepsilon,k)}$ in
$[\alpha - \varepsilon/2,\beta + \varepsilon/2]$ can be estimated
by
$$ C(\beta - \alpha + \varepsilon)R_n $$
and by the minimax principle and (\ref{eq:Reps}) the number of
eigenvalues of $\opr{H}_{R_{n(\varepsilon)}}$ in
$[\alpha,\,\beta]$ is estimated with the same relation. In this
way we have proved the bound
$$
\#\{j:\alpha < \lambda_j \leq \beta\} \leq C_\varepsilon +
C_0(\beta - \alpha + \varepsilon) R_n.$$
Finally, we substitute this result back to the right-hand side of
(\ref{eq:odhad}), and taking into account that $R_n$ can be chosen
arbitrarily large, we obtain the needed inequality
(\ref{eq:pur.22}) concluding thus the proof.
\end{proof}

\subsection*{Acknowledgments}

The research was supported by the Czech Academy of Sciences and
Ministry of Education, Youth and Sports within the projects
A100480501 and LC06002.

\end{document}